\label{sec:background}
\documentclass[runningheads]{llncs}
\usepackage{graphicx}
\usepackage{soul}
\usepackage{rotating}
\usepackage{graphicx}
\usepackage{amssymb}
\usepackage{blindtext}
\usepackage{makecell}
\usepackage{colortbl}
\usepackage{subcaption}
\usepackage{amsmath}
\usepackage[utf8]{inputenc}
\usepackage{csquotes}
\usepackage{cellspace}
\usepackage{array, tabularx, caption, boldline}
\usepackage{enumitem}
\usepackage{xcolor}
\usepackage{multirow}
\usepackage{booktabs}
\usepackage{bbding}

\usepackage{algorithm} 
\usepackage{algpseudocode} 

\usepackage{multirow,tabularx}
\usepackage{parskip}
\usepackage{subcaption}
\usepackage{caption}
\usepackage{varwidth}

\usepackage{bbm}

\usepackage{hyperref}

\usepackage{todonotes}

\begin{document}

\title{A simple LAD-LASSO coordinate descent algorithm for interactive browser-based GPU applications}
\titlerunning{LAD-LASSO}
\author{Stephen Wright\inst{1}\Envelope}
%
%
\institute{
  University of York, United Kingdom\\
  \Envelope \email{smw572@york.ac.uk} 
}

\maketitle

\begin{abstract}

Simultaneous variable selection and robust data fitting are important aspects of many mathematical modelling projects and a wide array of optimisation tools and techniques exist to support them. When the intention is to embed this capability in run-time interactive decision support tools running hundreds of such modelling tasks simultaneously on a GPU, the choices of implementation approach are more limited.

Recently, simple and fast Coordinate Descent algorithms have been proposed which can implement the LASSO approach to variable selection in conjunction with ordinary least squares (OLS) data fitting. However extending this to use the more robust Least Absolute Deviation (LAD) data fitting has been hampered by the multiple axis wise local minima that occur in the objective function for this LAD-LASSO approach.

This paper suggests that these multiple axis wise local minima form a locus which is monotonic in all the axes and that this locus has a convex objective function. Hence allowing the locus to be searched using a ternary chop algorithm that uses Coordinate Descent to identify multiple local minima (points on this locus) as required to find the global minimum.

The resulting algorithm is very simple making it practical to implement it as a single thread on a GPU. This opens up the possibility of running many hundreds of such threads in parallel using coarse parallelisation~\cite{buck2004toolkit}. These are early results in a wider project to explore the use of combinatorial sub sets of data in interactive mathematical modelling support frameworks.

\end{abstract}

\section{Introduction}
\label{Introduction}
 
When developing mathematical models for business decision support in areas such as the finance industry, agile interaction with intelligent non-specialists from a variety of backgrounds (and subject to diverse corporate IT policy rules) is often needed. This requires that any support environment involved is as agnostic as possible to the levels of internet connectivity and software approval and installation processes that may (or may not) be allowed. This makes client side calculation with the avoidance of client side SDK or library installation and/or configuration a design requirement for such tools. Hence this motivates a preference for Simple algorithms that can run locally in a browser. This is due to it's ubiquitous availability as well as the fact it can provide broad capabilities and runs in it's own pre-approved sandbox.

Of course the primary requirement remains that the parameter fitting is robust to outliers in the data and and is as explicable (i.e. as parsimonious) as possible. Least Absolute Deviation (LAD) regression is much more resistant to outliers than conventional OLS regression.~\cite{thanoon2015robust}  While the Least Absolute Shrinkage and Selection Operator (LASSO) helps to identify which explanatory variables are contributing most to the quality of the fitted model.~\cite{tibshirani1996regression} 

A combination of these two related approaches allows simultaneous variable selection and robust data fitting.~\cite{doi:10.1198/073500106000000251}~\cite{wang2006regularized} hence the interest in  the LAD-LASSO regularised linear regression method despite it's non smooth objective function introducing inconvenient computational considerations. It can be easily formulated as a Linear Programming (LP) problem ~\cite{vanderbei2015fast} although this can be computationally expensive
compared to say the Coordinate Descent LASSO-OLS method ~\cite{wright2015coordinate}.

These computational overheads in terms of speed and algorithmic complexity are particularly important when attempting to deliver interactive solutions with the type of fast update rate that this requires.They can also limit the potential to exploit massively parallel processing if the added complexity precludes an embarrassingly parallel solution. Recently, more general (non Coordinate Descent) algorithms have been proposed which allow LAD-LASSO searches however these come with the downside of requiring more complex code than the simple coordinate descent LASSO-OLS method ~\cite{shi2019descent1}~\cite{shi2019descent2}

Bringing these requirements together motivates this paper which describes a simple new coordinate descent approach for regularised, robust, linear (LAD-LASSO) parameter fitting. For development and demonstration purposes, this is implemented in Python in such a way as to avoid operations such as matrix inversion that would subsequently be inconvenient to program using the limited GPU capabilities (WebGL) that are directly available by default in most browsers. 

In this paper We have compared python versions of two new LAD-LASSO coordinate descent algorithms (One Ternary chop and one Quadrature) with a standard LP library (PuLP), and with a brute force (vertex evaluation) method. Early results have shown that on very small data sets both can be faster than PuLP. While on data sets that are more typical of the size of interest to our applications, (say 30 data points and up to 5 explanatory variables) the quadrature algorithm is slower than PuLP but can run on GPUs without requiring intermediate stream reduction operations which are computationally expensive on GPUs. 

A (static HTML) browser test framework is attached as supplementary material so that readers can inspect the code and test the algorithms locally on their own PC applied to data relevant to their own work. 

The rest of this paper is structured as follows: In the next Section~\ref{sec:background} We discuss the methods which underpin the work presented in this paper. Following that, the section ~\ref{sec:linear} on Linear Programming implementations of LAD-LASSO shows how LAD LASSO can be formulated as a special case of Linear Programming and solved either using a conventional (if heavyweight) Python Library or using a very simple brute force algorithm (that does not scale well) 

We then introduce in the  following section ~\ref{sec:CoordinateDescent} the topic of Coordinate Descent, it's application to smooth and non-smooth objective functions, and outline an algorithm which extends it's application to non smooth but convex objective functions such as LAD-LASSO. Finally the following section ~\ref{sec:ComparativePerformance} discusses Comparative performance of the four algorithms presented,  then provides conclusions and suggested further work ~\ref{sec:Conclusions}

\section{Background}
\label{sec:background}

In this section we present in more detail the theory behind the two techniques which we are merging to form a unified robust data fitting and variable selection algorithm.

\subsection{Least Absolute Deviations (LAD) Regression}
In an ideal world, the data input to a regression exercise will have a linear relationship between the dependent and explanatory variables with an additional known and "well behaved" noise distribution (such as a Gaussian) around this relationship. 

A data point may be considered an outlier if it does not fit this pattern and lies multiple standard deviations away from the mean. Unfortunately for standard least square linear regression a single arbitrarily large outlier can completely dominate the calculation.

The methods used to minimise this problem exploit the ability to vary the weighting applied to the residuals in the analysis.  e.g. in order of increasing robustness some example weighting schemes are….
\begin{itemize}
\item Sum of squared errors (Ordinary Least Squares or OLS) 
\item Sum of absolute deviations(Least Absolute Deviation or LAD)~\cite{thanoon2015robust}
\item Sum of squared error on fitted points with absolute deviation on outliers (e.g. Huber loss function)~\cite{huber1964robust}
\item Median (rank) or integer count methods ~\cite{zuo2021general}
~\cite{rosseeuw1987robust}
\end{itemize}
For our purposes in this paper we are focusing on the LAD approach, as it is conceptually simple and effective even if it is not computationally convenient (as it lacks an analytical solution). ~\cite{dietz1987comparison}

 Unfortunately when exploring a new data set, it is not just outliers that can cause us to draw misleading conclusions from our analysis. There can also be problems with multi-co-linearity, over-fitting and uninformative variables. 

These can all be mitigated by shrinking our estimated parameters to the smallest magnitude consistent with an adequate quality of fit. This is discussed in the next section on the LASSO technique. 

\subsection{The Least Absolute Shrinkage and Selection Operator (LASSO)}
Over-fitting in regression analysis occurs when the model ends up fitting the noise in the sample data rather than just the parameters of the underlying data generating process. This results in very good reported fit within the sample, but a high probability of a very poor fit out of sample data.

One cause of this is having too few data points for the number of parameters that you wish to fit. In the extreme case of having the same number of sample points as the number of parameters being fitted, the in sample fit can be exact!

Another cause is where there is poor variable selection and/or multi-co-linearity. (i.e. the variables being measured are not linearly independent of each other.) When this occurs, small changes in measured data values (or noise) can result in large changes in estimated parameter values. 

Parameter fitting can be seen as a type of optimisation problem where we seek to find the “best” parameter value to fit the underlying process.  It is characteristic of this type of optimisation that it is a diminishing returns effect, i.e. because the parameter search space is convex, a steepest descent search procedure will exploit the most significant search direction first. The least significant directions will be exploited last. 

Regularisation exploits this diminishing returns effect by adding a cost penalty to the magnitude of the fitted parameter values~\cite{tibshirani1996regression}. In this way, as the value of this cost penalty is varied, the steepest descent search procedure can be stopped early when the significance of any improvement in the in sample fit falls below the chosen benefit per unit parameter value change.

In the same way that robust regression methods vary by how they penalise the residuals in a data fitting exercise, regularisation methods vary in how they penalise the magnitude of the model parameters. The two main types of regularisation are Ridge Regression (which penalises the sum of the squares of the parameter values)~\cite{gruber2017improving} and LASSO regularisation (which penalises the sum of the absolute magnitudes of the parameter values)~\cite{tibshirani1996regression}.

Because the ridge regression parameter value is the square of the magnitude, it does not penalise small parameter values nearly as much as large ones. This favours models with many small parameter values.  We prefer the LASSO version of regularisation because it does not suffer from this problem, and so will tend to favour more parsimonious models (set more parameter weights to zero) than ridge regression.~\cite{andreas2016introduction} 

What LASSO does not do for you is to tell you the optimum value of the regularisation cost penalty. Hence Cross Validation is often used to provide guidance on the value of this parameter which gives the best balance between under-fitting and over-fitting tendencies in your model.~\cite{obuchi2016cross}

\section{Approach}
\label{sec:Approach}
Having discussed the robust data fitting and variable selection methods individually in Section~\ref{sec:background}, we now show how linear programming allows us to combine them in a simple unified algorithm in two different ways. Firstly using a Python linear Programming library and secondly by exploiting some special features of our LAD-LASSO task we show a very simple brute force vertex evaluation method. 

The benefit of having two options is that The standard python library is highly likely to give the correct result and is representative of current best practice and as such forms a good reference algorithm to test performance against.  The brute force method is guaranteed to give the correct answer and is so simple that it is difficult to set it up incorrectly (but is impractically slow on all but the smallest of problems). Hence it is a good cross check on having set up the LP library calls correctly. It also may be of interest in its own right for very small problems.

Having established the standard LP solutions to LAD-LASSO we next consider an innovative coordinate descent optimisation approach to find the minima of the LAD=LASSO objective function. The benefit of this is that these algorithms are more scalable than the brute force LP version discussed above and simpler to implement in parallel that the PuLP library routines.

\subsection{A unified LAD-LASSO Linear Programming algorithm}
\label{sec:linear}
The LASSO approach is often applied to OLS regression, however there is a natural symmetry between the LASSO and LAD approaches in that one penalises the absolute value of the residuals in a data fitting exercise, and the other penalises the absolute values of the parameter estimates. Because of this symmetry, the combined LAD-LASSO approach can be implemented as a simple unified Linear Programming (LP) algorithm.  

In particular, back to back slack and surplus variables can be added to each decision variable value as well as all the data points (constraints). This results in a standard form LP task that can be solved using one of the many available specialised LP libraries. 
\newpage
Or more formally,…….
The standard linear programming technique solves a problem in the following form……

We wish to minimise an objective function of the form
\begin{equation}
    f = cT \times x
\end{equation}
where $x$ is a vector of $d$ elements.

Given “d” variables:-
 \begin{center} 
             	x[1], . . . , x[d].
              
 	And “m” linear constraints in these variables
\begin{verbatim}
             	a[1] * x <= b[1], . . . , a[m]*x <= b[m]
              
 	         and Where       x[1] >= 0, . . . , x[d] >= 0.
    
\end{verbatim}

\end{center} 
We can new recast this formulation as follows …….
\begin{enumerate}
\item The vector of unknown parameter values $x$ can be replaced with a vector of length $2d$ comprising the non negative variables $x_p$ and $x_n$ (i.e. the x-positive and x-negative vectors respectively.)
\item To each constraint hyperplane equation in this parameter space we can add two non negative residual variables $r_p$ and $r_n$ (i.e. the residual-positive and residual-negative components respectively.)
\item We can then construct a cost function that weights the $x_p$ and $x_n$ vectors equally, and weights the $r_p$ and $r_n$ vectors at this weight times the regularisation constant selected.
\end{enumerate}
We now have a unified LP form of our parameter fitting exercise that can be solved by any of the standard LP libraries that are widely available. (In practice we have chosen to use the Python library PuLP.~\cite{mitchell2011pulp})

Having shown that we can cast the LAD-LASSO problem as a standard Linear Programming task, we can now exploit some of the special features of this formulation to allow a much simpler brute force version of the algorithm that is guaranteed to find the minimum value of the LAD-LASSO objective function but which is only useful for small problems as the computation time suffers a combinatorial explosion.

\subsection{A brute force version of the LP algorithm}
Because our augmented parameter search space is a Linear Program with a linear cost function we can guarantee that the optimum occurs at a vertex where “2d+2m” hyper-planes intersect. 

This means that in order to find the optimum we can evaluate the cost function at all the vertices and select the one exhibiting the minimum cost as the optimum.

When we do this we can also exploit a further simplifying factor that is unique to our LAD-LASSO LP task.  This simplification arises from the back to back formulation of our slack and surplus variables as this means that the objective function gradient only changes when one of the residuals (and/or one of the decision variable values) changes sign. 

In this special case of the LP problem, the vertex for the surplus variable becoming active and the vertex for the slack variable becoming inactive are coincident. Also these vertices all occur within the original decision variable space. Hence reducing the dimensionality of the space we need to search.

For large problems with a hundred or more constraints, this would still be a prohibitively large task, but for small problems, this brute force algorithm avoids the start up time required to initialise a standard LP library and exploits the unique simplifying structure of our problem.  

In practical tests on (very) small problems the brute force approach is much faster than PuLP However it also requires a matrix inversion which is inconvenient to implement on a GPU.

Note that…. 
\begin{itemize}
\item a) Because we are minimising the sum of all the pairs of  “xp” and ”xn” and of “rp” and “rn”  we can be sure that at all times one of each pair of slack and surplus variables will be exactly zero without needing to add an extra constraint to explicitly require this. 
\item  b) Also because these pairs of slack and surplus variables cover the full range of possible residual values, we can be sure that the full augmented parameter space is feasible so simplifying the task of finding an initial feasible solution.
\item  c) Because our augmented parameter search space is a Linear Program with a linear cost function we can guarantee that our minimisation exercise is a convex problem with a single global minimum.~\cite{dantzig2003linear}
\end{itemize}

\subsection{Coordinate Descent}
\label{sec:CoordinateDescent}
\subsection{Coordinate descent works well when finding the minimum of a smooth function But...}
Cyclical Coordinate Descent (CCD) is a simple, effective optimisation approach when applied to smooth convex functions. It works by choosing at each iteration in the descent, an element of the decision vector  and doing an objective function minimisation along this axis holding all the other members of the decision vector constant. The value of the selected decision vector element is then replaced by it's value at the calculated minimum and the process repeated for each decision vector element in turn until no more improvement can be found. This process is a surprisingly fast method for finding minima in smooth objective functions~\cite{gurbuzbalaban2017cyclic}. 

However when CCD is applied to a non-smooth objective function it can sometimes get stuck in local minima depending on how the contours align with the axes. For instance consider the contour plot below. At point $p$ (or more generally at any point on the dotted line passing through "p") the gradient of any direction parallel with the axes is increasing. As these are the only search directions allowed in a co-ordinate descent algorithm, the descent will terminate as soon as it arrives at any point on the dotted line.

\begin{figure}
\includegraphics[scale=0.8]{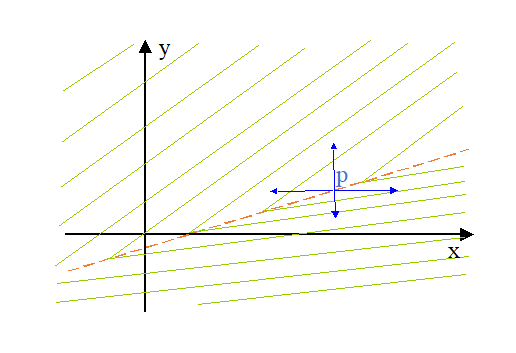}
\caption{Contour Plot Showing axis wise local minima\label{fig:contour}}
\end{figure}

Of course this is not a global minimum because as can be seen from the contours, a more general gradient descent algorithm would find a descent direction and proceed along the dotted line toward the right of the contour plot. However this comes at the considerable cost of having to search all possible directions for the next best step rather than just the coordinate directions.

\subsection{Some observations on local (axis wise) minima in otherwise convex functions}
It is quite possible for Cyclical Coordinate Descent to find the global minimum of a non smooth function if the discontinuities are aligned with the coordinate axes (as they are in the LASSO-OLS example). It may also be true that for a LAD problem, the discontinuities are sometimes aligned with the axes and sometimes not depending on the exact data values. However at this stage of the argument we will only consider the case of multiple locally axis wise minima in the objective function . Here we can make some observations as follows:- 
\begin{itemize}
\item Any vertical section through a convex function will have a singe minimum.
\item  If we take vertical sections aligned with each axis through an axis wise minimum, all the section wise minima will coincide at the axis wise minima
\item If this local minima is not a global minima, a descent path (not aligned with the coordinate axes) toward the global minima will pass through it.
\item This non aligned descent path will itself form a locus of further axis wise  minima.
\item The path of this locus will be monotonic in all it's coordinate values as otherwise it would need to recross a previous axis aligned section and create a new minimum in that section which proposition 1 does not allow. 
\item There is only one locus through the objective function space as otherwise there would need to be multiple minima in the individual vertical sections again this is precluded by proposition 1
\item The value of the objective function along this locus will itself be a convex function as it is a monotonic line drawn on a convex function.
\item If we perturb one coordinate of a local axis wise minima and use it as the starting point for a new cyclical coordinate descent we will end up at an adjacent local axis wise minima with a different objective function value.
\item It does not matter which coordinate we choose to perturb as by definition the axis wise local minima are not aligned with any axis.
\end{itemize}
From these observations we can conclude that the locus of axis wise local minima can itself be searched for a minimum value (i.e. the global minimum for the objective function) by using a technique such as a ternary search along one of the coordinate axes with each test point along this search line initiating a second stage search for the corresponding axis wise local minimum in the relevant sub space. Because we can calculate the piece-wise linear objective function for the original LAD-LASSO problem we can use a Binary, Ternary or Quadrature search for this second stage sub space minimisation as required. 

A discussion of some practical considerations of the resulting algorithm are provided in the next section.

\subsection{The Algorithm - Outline and practical considerations}
 The simplest version of the approach can be loosely summarised as:-
 \begin{figure}
     \includegraphics[width=\linewidth]{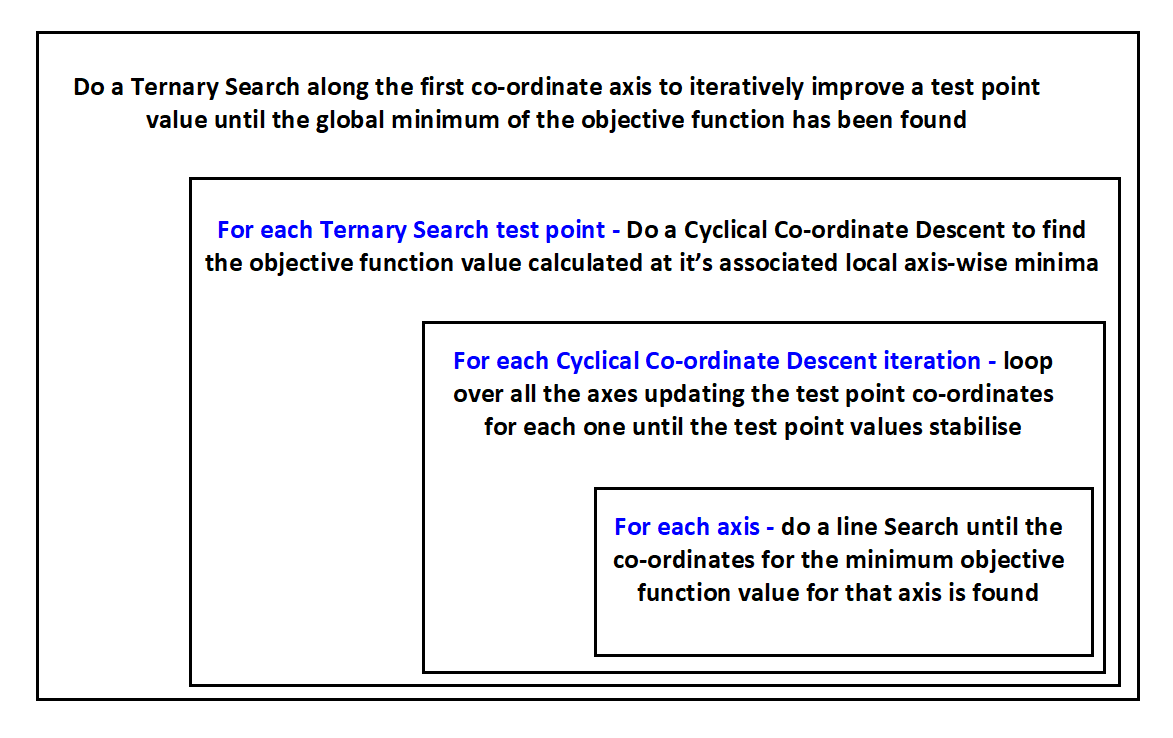}
     \caption{Coordinate Descent Algorithm Overview\label{fig:10point}}
 \end{figure}

Note that the locus of axis wise minima to be searched is illustrated by the dashed line in fig 1 above (for a 2D case)

Full details of this and the all the other algorithms discussed in this paper are provided in the attached supplementary material. However there are some practical details that need to be considered here viz:- 

In the previous section, we assumed that starting from a perturbed starting point on the axis chosen for the ternary search would always lead to a new locally linear axis wise minimum value. In practice, depending on the exact data set, the CCD algorithm might sometimes proceed to a global minima even when starting from multiple different points hence disrupting the ternary search. 

 In order to avoid this problem, we can restrict the sub searches spawned by the ternary search axis to preclude the CCD search stage varying the ternary search axis coordinate. 

 We also presumed that the initial search range contained the respective global and/or local minimums This might not always be the case. We can however test the initial search ranges provided and if the minimum does not lie in the provided range we can extend it until it does.

Finally, we have not mentioned that the LAD algorithm can give ambiguous results. This occurs when the piece-wise linear objective function has a minimum value segment which has a zero gradient. (This might arise when for instance data points are symmetrically arranged about an axis.)

We can accommodate this by having a (very small) minimum value for the regularisation parameter in order to disambiguate the objective function.

\section{Evaluation}\label{sec:ComparativePerformance}

 In order to evaluate the relative performance of our Coordinate Descent search algorithms we have compared them with that of a brute force (vertex evaluation) method and with that of a standard Python LP library (PuLP) using data sets generated  using Scikit-learn’s "Make Regression" synthetic data set generator. 
 
 All four methods can give the same results to the selected accuracy so we have compared the execution time and how it varies over a range of one to five explanatory variables and between 10 and 30 data points.

 \begin{figure}
     \includegraphics[width=1.0\linewidth]{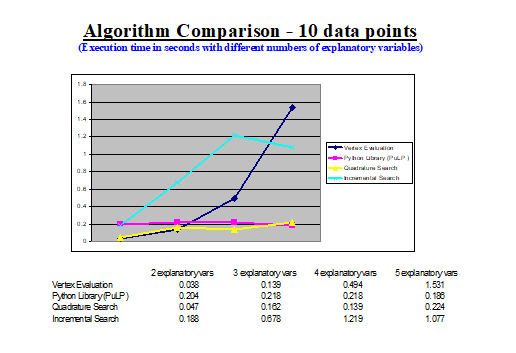}
     \caption{Relative run time with 10 data points}\label{fig:10point}
 \end{figure}
 
 \begin{figure}
     \includegraphics[width=\linewidth]{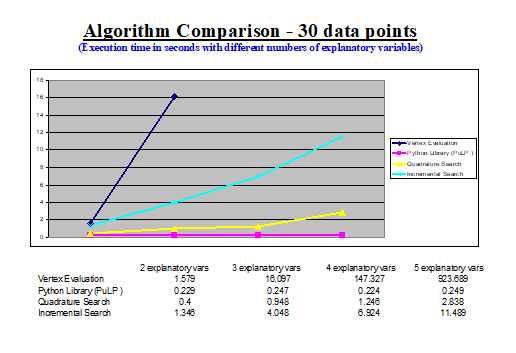}
     \caption{Relative run time with 30 data points}\label{fig:30point}
 \end{figure}

Please note that the figures represented in Figures~\ref{fig:10point} are indicative, the relative performance achieved in practice will vary with the exact data set, the choice of precision required and many other detailed implementation decisions.

\section{Conclusions and suggested further work} 
\label{sec:Conclusions}
This paper has presented a simple approach that allows coordinate descent to be applied to the LAD-LASSO simultaneous variable selection and robust data fitting task despite the non smooth objective function involved. The quadrature algorithm in particular has similar performance relative to a standard LP library implemented in Python (i.e. PuLP) when used on very small low dimension data sets. It is also expected to be more scaleable (and easier to implement on a GPU) than the brute force vertex evaluation algorithm we presented for comparison purposes. 

The main limitation of this coordinate descent approach is that  there is a speed versus accuracy trade off where the most appropriate choice of the hyper-parameter values controlling this trade off are context specific. Hence users may need to do some hyper-parameter tuning in order to get the best performance from the method.

However, these are early results hence the following  further work is suggested :-  
\begin{itemize}
\item We need to check the logic in the section "some observations on local (axis wise) minima in otherwise
convex functions" to confirm that the intuition is correct and if so, then prove them in a more formal mathematical manner
\item we need to test convergence on a wide variety of data fitting problems in addition to the small number of simple  test cases used so far
\item we need to do an algorithm complexity analysis to describe how the performance of these algorithms scale.
\item we need to add automatic cross validation to choose the best value of the regularisation parameter
\item we need to test the motivating assumptions that regularising a combinatorial array of multiple small sub sets of data points will reliably identify non-informative variables, over-fitting etc.
\item We need to implement the WebGL version of one or both of the LAD-LASSO algorithms discussed here to confirm that the single thread Browser-GPU implementation has similar performance to the Python implementation. This might then be extended to demonstrate coarse parallelisation by perhaps calculating multiple regularisation settings while simultaneously  cross validating the analysis to choose the preferred value of these options.

\end{itemize}
\bibliographystyle{splncs04}
\bibliography{refs}

\begin{thebibliography}{10}
\providecommand{\url}[1]{\texttt{#1}}
\providecommand{\urlprefix}{URL }
\providecommand{\doi}[1]{https://doi.org/#1}

\bibitem{andreas2016introduction}
Andreas, C., Guido, S.: Introduction to Machine Learning with Python: A Guide
  for Data Scientists. O'Reilly Media, Incorporated (2016)

\bibitem{buck2004toolkit}
Buck, I., Purcell, T.: A toolkit for computation on gpus, gpu gems (2004)

\bibitem{dantzig2003linear}
Dantzig, G.B., Thapa, M.N.: Linear programming: Theory and extensions, vol.~2.
  Springer (2003)

\bibitem{dietz1987comparison}
Dietz, E.J.: A comparison of robust estimators in simple linear regression.
  Tech. rep., North Carolina State University. Dept. of Statistics (1987)

\bibitem{gruber2017improving}
Gruber, M.: Improving efficiency by shrinkage: The James--Stein and Ridge
  regression estimators. Routledge (2017)

\bibitem{gurbuzbalaban2017cyclic}
Gurbuzbalaban, M., Ozdaglar, A., Parrilo, P.A., Vanli, N.: When cyclic
  coordinate descent outperforms randomized coordinate descent. Advances in
  Neural Information Processing Systems  \textbf{30} (2017)

\bibitem{huber1964robust}
Huber, P.J.: Robust estimation of a location parameter: Annals mathematics
  statistics, 35. Ji, S., Xue, Y. and Carin, L.(2008),‘Bayesian compressive
  sensing’, IEEE Transactions on signal processing  \textbf{56}(6),
  2346--2356 (1964)

\bibitem{mitchell2011pulp}
Mitchell, S., OSullivan, M., Dunning, I.: Pulp: a linear programming toolkit
  for python. The University of Auckland, Auckland, New Zealand  \textbf{65}
  (2011)

\bibitem{obuchi2016cross}
Obuchi, T., Kabashima, Y.: Cross validation in lasso and its acceleration.
  Journal of Statistical Mechanics: Theory and Experiment  \textbf{2016}(5),
  053304 (2016)

\bibitem{rosseeuw1987robust}
Rosseeuw, P., Leroy, A.: Robust regression and outlier detection, john willey
  \& sons. Inc., New York  (1987)

\bibitem{shi2019descent1}
Shi, Y., Feng, Z., Yiu, K.F.C.: A descent method for least absolute deviation
  lasso problems. Optimization Letters  \textbf{13},  543--559 (2019)

\bibitem{shi2019descent2}
Shi, Y., Ng, C.T., Feng, Z., Yiu, K.F.C.: A descent algorithm for constrained
  lad-lasso estimation with applications in portfolio selection. Journal of
  Applied Statistics  \textbf{46}(11),  1988--2009 (2019)

\bibitem{thanoon2015robust}
Thanoon, F.H.: Robust regression by least absolute deviations method.
  International Journal of Statistics and Applications  \textbf{5}(3),
  109--112 (2015)

\bibitem{tibshirani1996regression}
Tibshirani, R.: Regression shrinkage and selection via the lasso. Journal of
  the Royal Statistical Society: Series B (Methodological)  \textbf{58}(1),
  267--288 (1996)

\bibitem{vanderbei2015fast}
Vanderbei, R.J.: Fast algorithms for lad lasso problems  (2015)

\bibitem{doi:10.1198/073500106000000251}
Wang, H., Li, G., Jiang, G.: Robust regression shrinkage and consistent
  variable selection through the lad-lasso. Journal of Business \& Economic
  Statistics  \textbf{25}(3),  347--355 (2007).
  \doi{10.1198/073500106000000251},
  \url{https://doi.org/10.1198/073500106000000251}

\bibitem{wang2006regularized}
Wang, L., Gordon, M.D., Zhu, J.: Regularized least absolute deviations
  regression and an efficient algorithm for parameter tuning. In: Sixth
  International Conference on Data Mining (ICDM'06). pp. 690--700. IEEE (2006)

\bibitem{wright2015coordinate}
Wright, S.J.: Coordinate descent algorithms. Mathematical programming
  \textbf{151}(1),  3--34 (2015)

\bibitem{zuo2021general}
Zuo, Y.: On general notions of depth for regression  (2021)

\end{thebibliography}

\end{document}